\documentclass[a4paper,fleqn,usenatbib]{mnras}

\usepackage{newtxtext,newtxmath}
\usepackage{upgreek}
\usepackage[T1]{fontenc}
\usepackage{ae,aecompl}
\usepackage{caption}
\usepackage{subcaption}
\usepackage{multirow}

\usepackage{graphicx}	% Including figure files
\usepackage{amsmath}	% Advanced maths commands
\usepackage{upgreek}
\binoppenalty=\maxdimen
\relpenalty=\maxdimen

\usepackage{xcolor} % Required for specifying colors by name
\usepackage[colorinlistoftodos]{todonotes}
\usepackage{ulem}

\graphicspath{{./fig/}}

\title[FRB 121102 and its environment]{FRB 121102: drastic changes in the burst polarization contrasts with the stability of the persistent emission}
\author[A. Plavin et al.]{
A. Plavin,$^{1}$\thanks{E-mail: alexander@plav.in}
Z. Paragi,$^{2}$
B. Marcote,$^{2}$
A. Keimpema,$^{2}$
J. W. T. Hessels,$^{3,4}$
K. Nimmo,$^{3,4}$
\newauthor
H. K. Vedantham,$^{3}$
L. G. Spitler,$^{5}$
\\
$^{1}$Astro Space Center of Lebedev Physical Institute, Profsoyuznaya 84/32, 117997 Moscow, Russia\\
$^{2}$Joint Institute for VLBI ERIC (JIVE), Oude Hoogeveensedijk 4, 7991 PD Dwingeloo, The Netherlands\\
$^{3}$ASTRON, The Netherlands Institute for Radio Astronomy, Oude Hogeveensedijk 4, 7991PD, Dwingeloo, The Netherlands\\
$^{4}$Anton Pannekoek Institute for Astronomy, University of Amsterdam, Science Park 904, 1098 XH Amsterdam, The Netherlands\\
$^{5}$Max-Planck-Institut f\"ur Radioastronomie, Auf dem H\"ugel 69, D-53121 Bonn, Germany
}

\date{Accepted 18 February 2022. Received 18 February 2022; in original form 25 October 2021}

\pubyear{2022}

\begin{document}
\label{firstpage}
\pagerange{\pageref{firstpage}--\pageref{lastpage}}
\maketitle

\begin{abstract}

We study milliarcsecond-scale properties of the persistent radio counterpart to FRB~121102 and investigate the spectro-polarimetric properties of a bright burst. For the former, we use European VLBI Network (EVN) observations in 2017 at 1.7 and 4.8\,GHz. For the latter, we re-analyse the 1.7-GHz data from the 100-m Effelseberg telescope taken in 2016. These observations predate other polarimetric studies of FRB~121102, and yield the highest burst Faraday rotation measure (RM) to date, $\mathrm{RM} = 1.27\cdot10^5$~rad~m$^{-2}$, consistent with the decreasing RM trend. The fractional polarization of the burst emission is 15\% at 1.7\,GHz. This can be reconciled with the high fractional polarization at higher frequencies if the Faraday width of the burst environment is 150~rad\,m$^{-2}$ -- a bare 0.1\% of the total Faraday rotation. The width may originate from minor non-uniformities in the Faraday screen, or from effects in the emitting region itself. The upper limit on the persistent source size is 1\,pc, barely consistent with a young supernova (SN) scenario. The flux variability limit of $<10\%$ is not in favor of the young SN scenario, and challenges other interpretations as well. The fractional polarization of the faint persistent source is constrained at $<25\%$ at 4.8\,GHz ruling out a common origin with the highly polarized individual bursts.

\end{abstract}


%%%%%%%%%%%%%%%%%%%%%%%%%%%%%%%%%%%%%%%%%%%%%%%%%%

%%%%%%%%%%%%%%%%% BODY OF PAPER %%%%%%%%%%%%%%%%%%

\section{Introduction}

Fast radio bursts (FRBs) are millisecond-duration radio transients that show high dispersion measures (DMs), consistent with an extragalactic origin \citep{lorimer2007,petroff2019}. They were initially detected using the 64-m radio telescope at the Parkes Observatory at frequencies $\sim 1$\,GHz, and later also by a number of other single-dish radio telescopes or interferometers \citep[e.g.][]{spitler2014, caleb2017, shannon2018, amiri2019, ravi2019}. For a full list of FRBs detected so far, see the FRB Catalogue\footnote{\url{http://www.frbcat.org}} \citep{petroff2016}. Single-dish observations allowed localisations of several arcminutes at best, and therefore the host galaxies of FRBs could not be unambiguously identified at first. 
Assuming that the observed DMs arise predominantly in the intergalactic medium, the derived distances of $\sim 0.1-10$\,Gpc imply isotropic luminosities on the order of  $10^{38-43}$~erg\,s$^{-1}$.  In the past decade there have been a great number of theories developed to explain the phenomenon \citep[e.g.][]{katz2018, 2019PhR...821....1P}, many  of which invoke neutron stars. However, at the time neutron stars were not known to produce FRB-like luminosities, and other types of Galactic sources were also hypothesised that could naturally account for the observed DM \citep[e.g.][]{loeb2014} -- but these models also appeared to suffer from inconsistencies.

Confirming the extragalactic nature of FRBs was aided by the discovery of the first repeating source, FRB~121102 \citep{spitler2016}, which was found using the 305-m Arecibo telescope. A repeating source made it possible to organise interferometric follow-up observations, and led to precise localisation with the Karl G. Jansky Very Large Array (VLA) on sub-arcsecond scales \citep{chatterjee2017}.  Its position was further refined with the very long baseline interferometry (VLBI) technique to milliarcsecond scales using the European VLBI Network \citep[EVN;][]{marcote2017}. This was aided by the real-time correlation electronic VLBI (e-VLBI for short) capability of the EVN, which offers flexibility to one of the most-sensitive VLBI arrays to study transient phenomena at the highest-possible angular resolution; short-transient detection has been an important driver to the e-VLBI developments \citep{paragi2016}. Since this important discovery, sub-arcsecond localizations of FRBs have been achieved in a number of cases \citep[e.g.][]{bannister2019, prochaska2019, ravi2019, macquart2020}, including those from single-pulse bursts (non-repeaters). But milliarcsecond scales have only been probed for FRB~121102 \citep{marcote2017} and FRB~180916.J0158+65 \citep{marcote2020} with the EVN, and recenly for both FRB~20201124a and FRB~20200120E in M81, with an ad-hoc array of EVN telescopes in the PRECISE program \citep{marcote2021,kirsten2021}.

FRB~121102 lies within a star-forming region of a metal-poor dwarf galaxy at a redshift of 0.19273(8) \citep{tendulkar2017,bassa2017}. Curiously, there is a persistent radio source co-located with the burst source. Its VLBI and multi-band properties are in agreement with either a low-luminosity active galactic nucleus (AGN; though this source is not at the optical centre of the amorphous dwarf host galaxy), or a young magnetar nebula \citep{marcote2017,scholz2017}. The  detection of bursts at higher radio frequencies ($\sim 4-8$\,GHz) using Arecibo \citep{michilli2018} and the Robert C. Byrd Green Bank Telescope \citep[GBT;][]{gajjar2018} revealed a very high rotation measure RM = $10^5$\,rad\,m$^{-2}$ that was seen to decrease by $\sim 10$\% within 7 months \citep{michilli2018,gajjar2018}.  The extreme and variable magneto-ionic environment suggests that the source of the bursts may be a magnetar orbiting a massive black hole, or that the high RM originates in a highly-magnetized wind nebula or a young supernova remnant surrounding a neutron star \citep{michilli2018}. Further studies of the milliarcsecond properties and its RM dependence on time- and frequency are required to constrain the various models better.

The earlier VLBI results on FRB~121102 were achieved by repeated observing campaigns throughout its active periods in 2016 at a centre frequency of 1.7\,GHz, using an array of western EVN telescopes and Arecibo. This resulted in great sensitivity, but limited resolution and a very elongated restoring beam, which is not ideal for astrometry. 
In this paper, we report on continued EVN monitoring of FRB~121102 at two frequencies, aiming at a more balanced $uv$-plane coverage and a higher resolution compared to previous work. These observations provide an improved astrometry for the persistent radio counterpart, a robust measurement of its spectral index, and the first constraints on its fractional polarization considering different RM values. In addition, we analyse the polarization properties of the brightest EVN-detected burst at 1.7\,GHz \citep[previously reported by][]{marcote2017} and provide an estimate of its RM. This is the earliest epoch for which an RM has been determined for FRB~121102 as well as the lowest radio frequency at which polarization of this source has been detected.

In \autoref{s:obs} we present the observations and describe the basic data processing. Next we present the observational results in \autoref{s:results} and discuss their interpretation in \autoref{s:discuss}.

\section{Observations and Data Reduction}
\label{s:obs}

Earlier 2016 EVN observations of FRB~121102 were presented in \citet{marcote2017}. We reanalyse the brightest burst caught in that project at 1.7~GHz on 2016 September 20, utilizing  data from the Effelsberg telescope. These observations employed the VLBI backend at Effelsberg instead of the pulsar backend typical for single-dish transient studies. The VLBI backend provides a much higher spectral and temporal resolution and allows us to study the burst's polarization properties.

We also performed dedicated FRB~121102 observations with the EVN at 1.7 and 4.8\,GHz in three 12-h observing sessions from February to November 2017. These observations included the 305-m William E. Gordon Telescope at the Arecibo Observatory, Effelsberg, Lovell Telescope or Mk2 at Jodrell Bank, Medicina, Noto, Onsala, Yebes, Hartebeesthoek, Toru\'n, Westerbork single dish, Tianma, Urumqi, Svetloe, Zelenchukskaya, Badary, Irbene, Green Bank, and Robledo. The full bandwidth was divided in eight sub bands (also called \textit{intermediate frequencies}, IFs) of 16\,MHz with a time integration of 2\,s and full polarization.

Our 2017 EVN observations caught no bursts from FRB~121102, and we use this dataset to analyze the persistent radio counterpart. These dedicated observations included more telescopes than previous studies \citep[e.g.][]{marcote2017}. A better coverage of spatial frequencies and a higher sensitivity allows us to study the counterpart in more detail.

\subsection{Effelsberg Single-dish Data}
\label{s:method_sd}

Coherently de-dispersed high time- and spectral resolution auto-correlations were produced for the brightest burst from the 2016 EVN localization of FRB~121102~\citep{marcote2017},
which had a duration of $\sim 2$\,ms and a peak flux density of $\sim 11$\,Jy. The data were correlated using the SFXC software correlator~\citep{keimpema2015} outputting all four
polarization products using 3.9-kHz spectral channels and 256-$\upmu$s integration times. There are eight 16-MHz IFs that are recorded and processed separately; this results in 8\,192 spectral channels per IF and 65\,536 channels in total.

The amplitudes of both polarizations were calibrated independently; to this end we used the 80-Hz continuous calibration signal injected into the data by the receiver backend. By measuring the total power in the parallel hands, the system temperature $T_\mathrm{sys}$ can be accurately estimated \citep{vlbamemo34}. Thus, we assume the statistical amplitude uncertainty equal to the random scatter of $T_\mathrm{sys}$ within 1 minute around the burst; a 10\% systematic uncertainty is added, typical for VLBI observations. We flag (i) $4\%$ of channels at both edges of each IF as the bandpass effect is evident there, and (ii) channels with off-pulse noise level more than two times higher than the average. Together this constitutes $11\%$ of all channels.

Given previous results \citep{michilli2018}, we expected  potential linear polarization at a high rotation measure, and possibly also circular polarization at a high conversion measure as well \citep{2019MNRAS.485L..78V,2019ApJ...876...74G}. Thus, the Stokes parameters cannot be simply averaged across the whole band or even across 16~MHz IFs. We also cannot employ the typically used method --- measuring the electric vector position angle (EVPA) independently in each frequency channel --- because of the very low signal-to-noise ratios (S/N) for individual channels. The S/N is low compared to 5-GHz studies of FRB~121102 due to several reasons: (i) the Faraday rotation is proportional to $\lambda^2$, so in the 1.7-GHz band channels need to be about three times narrower to be able to detect the same rotation measure; (ii) the most sensitive observations of this burst were performed at Arecibo, but they could not be used for polarization studies because of a very strong leakage, estimated to be about $40\%$; (iii) the linearly polarized signal itself turns out to be much weaker than at higher frequencies, as we show later in \autoref{s:results}.

Instead of independent polarization angle measurements within each channel, we perform Fourier transforms of the Stokes quantities in the $\lambda^2$ domain. This is basically the ``RM synthesis'' approach described in \cite{2005A&A...441.1217B}: the Fourier transform of Stokes $Q$ and  $U$ yields estimates of the linearly polarized flux $S(\mathrm{RM})$ for different rotation measures RM. Similarly, transforming Stokes $V$ allows measuring circularly polarized flux $S(\mathrm{CM})$ for different linear-to-circular conversion measures CM. In the following, we refer to $S(\mathrm{RM})$ and $S(\mathrm{CM})$ as the rotation measure spectrum and conversion measure spectrum, respectively. These spectra are computed as follows:
$$S'(\mathrm{RM}) = \frac{1}{n} \sum_{j=1}^n [Q(\lambda_j) + \mathrm{i} U(\lambda_j)] \mathrm{e}^{-2\mathrm{i} \cdot \mathrm{RM} \lambda_j^2} = \frac{1}{n} \sum_{j=1}^n \mathrm{RL}(\lambda_j) \mathrm{e}^{-2\mathrm{i} \cdot \mathrm{RM} \lambda_j^2}$$
and
$$S'(\mathrm{CM}) = \frac{1}{n} \sum_{j=1}^n V(\lambda_j) \mathrm{e}^{-2\mathrm{i} \cdot \mathrm{CM} \lambda_j^2} = \frac{1}{n} \sum_{j=1}^n \frac{\mathrm{RR}(\lambda_j) - \mathrm{LL}(\lambda_j)}{2} \mathrm{e}^{-2\mathrm{i} \cdot \mathrm{CM} \lambda_j^2}.$$
Here $\mathrm{RR}(\lambda), \mathrm{LL}(\lambda)$ and $\mathrm{RL}(\lambda)$ are the measured parallel and cross-hand spectra. They are dominated by noise in each channel individually, but the Fourier transform effectively averages this noise out. The absolute values of the resulting spectra --- $|S'(\mathrm{RM})|$ and $|S'(\mathrm{CM})|$ --- represent the flux density at rotation measure RM and conversion measure CM; the phase $\frac{1}{2} \mathrm{Arg}\{S'(\mathrm{RM})\}$ is the EVPA of the flux linearly polarized at RM normalized to $\lambda = 0$~m. Computed $|S'(\mathrm{RM})|$ and $|S'(\mathrm{CM})|$ are positively biased as polarized flux measurements; we follow \cite{2001ApJ...553..341E} and de-bias these values by setting $|S|=\sqrt{|S'|^2 - \sigma^2}$ when $|S'| > 1.5\sigma$, and $S=0$ otherwise. Here $\sigma$ is the noise level in $S$.

Note that the computed profiles trivially include linear and circular polarization that are constant across the band: it corresponds to $\mathrm{RM} = 0$~rad~m$^{-2}$ and $\mathrm{CM} = 0$~rad~m$^{-2}$. Our further analysis is based only on the absolute values $|S(\mathrm{RM})|$ and $|S(\mathrm{CM})|$ of these complex quantities. The phase $\mathrm{Arg}\{S(\mathrm{RM})\}$ is dependent on the calibration of the cross-polarization delay and the EVPA. This calibration could not be performed for our experiment: adequate calibrators were not observed. We do not perform any ionospheric correction: RM values of interest are orders of magnitude higher than the ionospheric contribution, which does not exceed tens of rad~m$^{-2}$ \citep{malins2018}. Based on \cite{2005A&A...441.1217B}, we estimate that our analysis is sensitive to Faraday rotation up to $2\cdot10^7$\,rad\,m$^{-2}$ at a 50\% level --- much further beyond the range of interest at $\approx 10^5$\,rad\,m$^{-2}$. The effective resolution in the rotation measure space is defined by the spread function width: $\mathrm{FWHM} \approx 5000$\,rad\,m$^{-2}$ for each of the eight individual IFs, and $\mathrm{FWHM} \approx 2500$\,rad\,m$^{-2}$ for a pair of consecutive IFs. The accuracy of the peak location is significantly better than these FWHM values.

As demonstrated later, we see no evidence of multi-peak structure in rotation measure profiles. This justifies using the maximum absolute value $|S(\mathrm{RM})|$ as the estimate of linearly polarized flux and its rotation measure. This approach is exactly equivalent to a least-squares fitting of $\mathrm{RL}(\lambda_j) = S_{\text{Lin}} \mathrm{e}^{2\mathrm{i} \cdot \mathrm{RM} \lambda_j^2}$ with $\mathrm{RM}$ and $S_{\text{Lin}}$ as free parameters. We estimate statistical uncertainties of individual $S(\mathrm{RM})$ values and of their maxima using a non-parametric bootstrap with 1000 iterations. Each iteration consists of repeating the same computation for a random subset of individual channels.

The polarization analysis is first performed separately for each of the eight 16-MHz IFs because instrumental polarization differences between them are not calibrated. However, there are only four independent hardware bands that correspond to pairs of consecutive IFs. Both IFs within a single hardware band undergo the same transformations and are influenced by the same instrumental effects. Thus, it is possible to combine IFs into pairs to increase the sensitivity. In \autoref{s:results}, we evaluate and compare both approaches. We perform an additional check by calculating polarization properties twice: for the burst peak, and time-integrated over the whole burst. This is useful to ensure consistency, even though we find our observations not sensitive enough to study temporal variations across the burst.

The burst observations were performed with circularly polarized feeds, and circular polarization measurements are strongly affected by residual amplitude errors. In contrast, linear polarization measurements $S(\mathrm{RM})$ are based on cross-hand $\mathrm{RL}(\lambda)$ spectra only, and to the first order are free from amplitude miscalibrations. The D-terms or cross-polarization leakages were estimated to be below $3\%$ based on the off-burst noise properties, and were not calibrated for. These leakages have no first-order effect on parallel-hand spectra $\mathrm{RR}(\lambda), \mathrm{LL}(\lambda)$ and thus on $S(\mathrm{CM})$ estimates. For strongly linearly polarized sources, this would not strictly true because the $\mathrm{LR}\times D^{R*}$ and the $\mathrm{RL}\times D^{L*}$ terms become non-negligible even at the first order \citep[cf.][]{Paragi2004}; as we show later, this is not the case at the frequencies we probe. All these instrumental effects vary slowly with frequency: on the scale of IF bandwidth (16~MHz) or larger. Thus, they can only lead to spurious signal in $S(\mathrm{RM})$ and $S(\mathrm{CM})$ at $\mathrm{RM}, \mathrm{CM} \lesssim 3000$~rad m$^{-2}$. As we show later in \autoref{s:results}, no linear polarization signal is detected in this RM region anyway. Circular polarization estimates close to $\mathrm{CM} = 0$, however, remain limited by these calibration uncertainties.

\subsection{EVN Interferometric Data}
\label{s:method_vlbi}

We used J0529+3209, which is $1\degr$ away from FRB~121102, as the phase calibrator in all sessions. Phase-referencing cycles were scheduled with 3.5\,min on the target and 1.5,min on the phase calibrator. In addition we had an in-beam (inside the primary beam) check source 1.8\,arcmin away from the target --- the so-called VLA2 source in \citet{marcote2017} --- that is compact and has a flux density of $\sim 2\,\ \mathrm{mJy}$, an order of magnitude brighter than the target source. The target/check source separation lies well within the primary beam of all EVN stations except Arecibo at $4.8$\,GHz. We use this source to evaluate if a significant fraction of flux is lost when doing phase-referencing without additional self-calibration of the target, and to provide more accurate relative astrometry estimates including proper motion.
Checking for proper motion in either the persistent radio source and/or the source of bursts was part of the original goals of these follow-up observations to completely rule out even the tiniest possibility for chance coincidence alignment of a Galactic source with the host dwarf galaxy of FRB~121102 reported following its discovery \citep{chatterjee2017,marcote2017,tendulkar2017,bassa2017}.

The correlated data were calibrated using standard VLBI procedures using {\tt AIPS}\footnote{The Astronomical Image Processing System (AIPS) is a software package produced and maintained by the National Radio Astronomy Observatory (NRAO).} \citep{greisen2003} and {\tt ParselTongue} \citep{kettenis2006}, including {\it a-priori} amplitude calibration from the EVN Pipeline, as described in the EVN Data Reduction Guide\footnote{\url{https://www.evlbi.org/evn-data-reduction-guide}.}. The phases were corrected by fringe-fitting the calibrator J0529+3209. This source was then imaged and self-calibrated, and the derived amplitude and phase corrections were applied to the target and in-beam check source, which were finally imaged. We apply primary beam corrections based on a Gaussian model for the different antennas when the target phase center was different from the telescope pointing position. Cross-hand polarization delays were removed using the procedure described in {\tt AIPS} documentation using the data for J0237+2848, and the antenna feed parameters were determined using the LPCAL task with J0518+3306 as the calibrator. We then fit a model consisting of a single circular Gaussian component in {\tt Difmap} \citep{shepherd1994} for both the target and check source to calculate their positions, flux densities, and sizes. We estimate the linearly polarized flux using {\tt AIPS} in two different ways. The first is done by averaging over the whole band, which corresponds to constant polarization assuming RM$\sim0$~rad~m$^{-2}$ along the line of sight of the persistent radio source. The other includes scanning over $4000$ trial RMs in the range $\pm2\cdot10^5$~rad~m$^{-2}$ (i.e. including in the range the RM of single bursts detected in FRB~121102), computing dirty images for each of them.

We consider the source as not detected when no pixel in the dirty image within 50\,mas of the target position has a value above the $5\sigma$ rms noise level. This is the case for polarized emission only, and we take this $5\sigma$ level of the dirty image as an upper limit in this case. We take the synthesized beam size divided by S/N as the formal error of our position estimates, and the dirty map noise level as the formal error of flux density estimates. We also account for amplitude calibration uncertainty by adding 10\% to the flux density error, which is the typically assumed systematic amplitude error for VLBI observations.

We tried using self-calibration on the $\approx2$\,mJy in-beam check source to improve the solution for our target, which is an order of magnitude fainter. However, applying the solutions derived from self-calibration did not help, and actually lead to worse coherence. This likely happens because the check source is relatively faint, and its self-calibration solution includes a significant noise component.

\section{Results}
\label{s:results}

\subsection{Burst emission}
\label{s:res_burst}

\begin{figure}
    \centering
    \includegraphics[width=\linewidth]{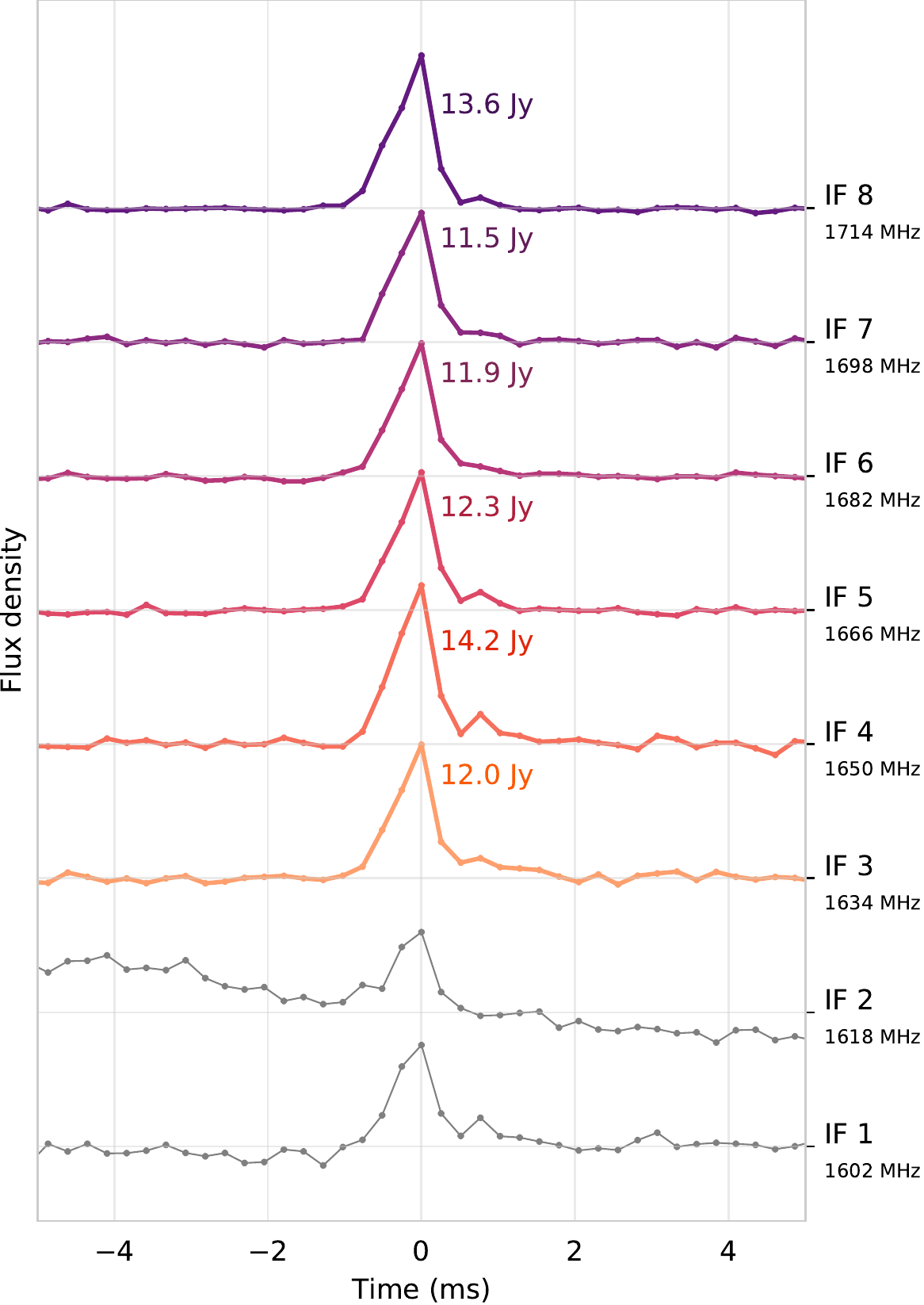}
	\caption{The temporal profile of the FRB~121102 burst on 2016 Sep 20 as seen by Effelsberg. Frequency subbands (IFs) are vertically separated for clarity. Time is shown relative to the peak of the emission that was simultaneous in all subbands.}
	\label{f:burst_flux}
\end{figure}

\begin{figure}
	\includegraphics[width=\linewidth]{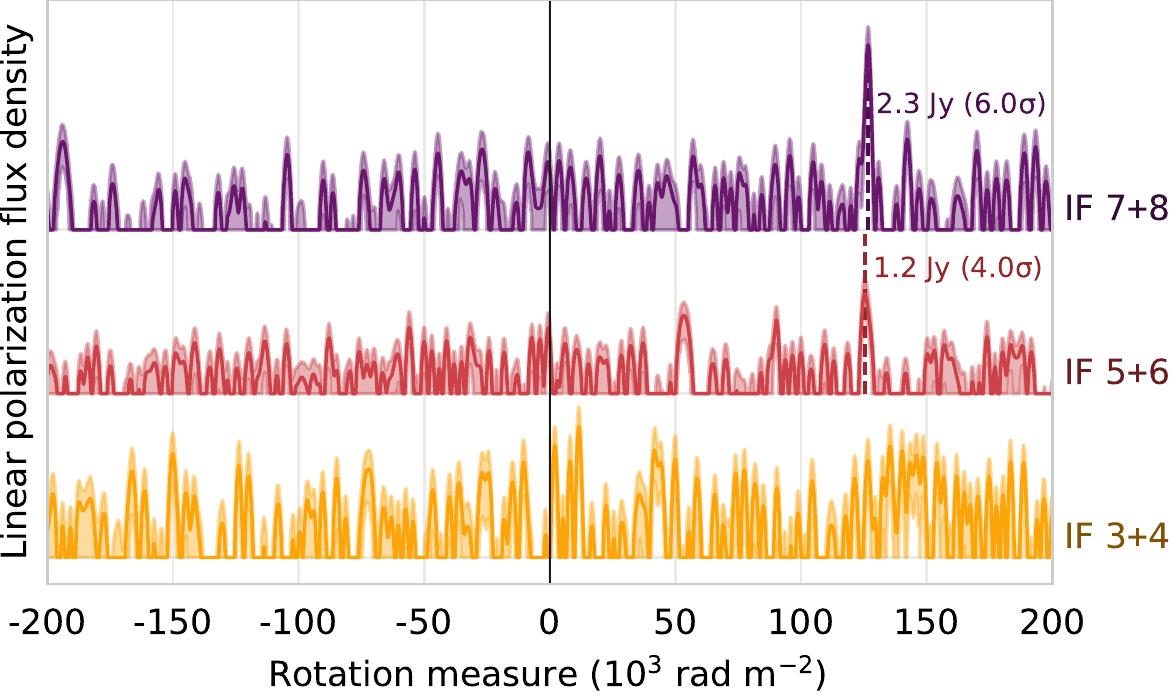}
	\caption{Faraday rotation profiles of the burst, see \autoref{s:results} for discussion. All measurements are in the observer frame of reference.
	Profiles correspond to the burst peak, $t=0$~ms in \autoref{f:burst_flux}. Shaded areas represent statistical pointwise 68\% confidence intervals. The three hardware bands of the telescope are shown individually. All peaks above the $3.5\sigma$ level are labelled. See Appendix~\ref{a:rm_profiles} for profiles split into individual IFs or integrated over the whole burst duration.}
	\label{f:rm_profile}
\end{figure}

\begin{figure}
    \includegraphics[width=\linewidth]{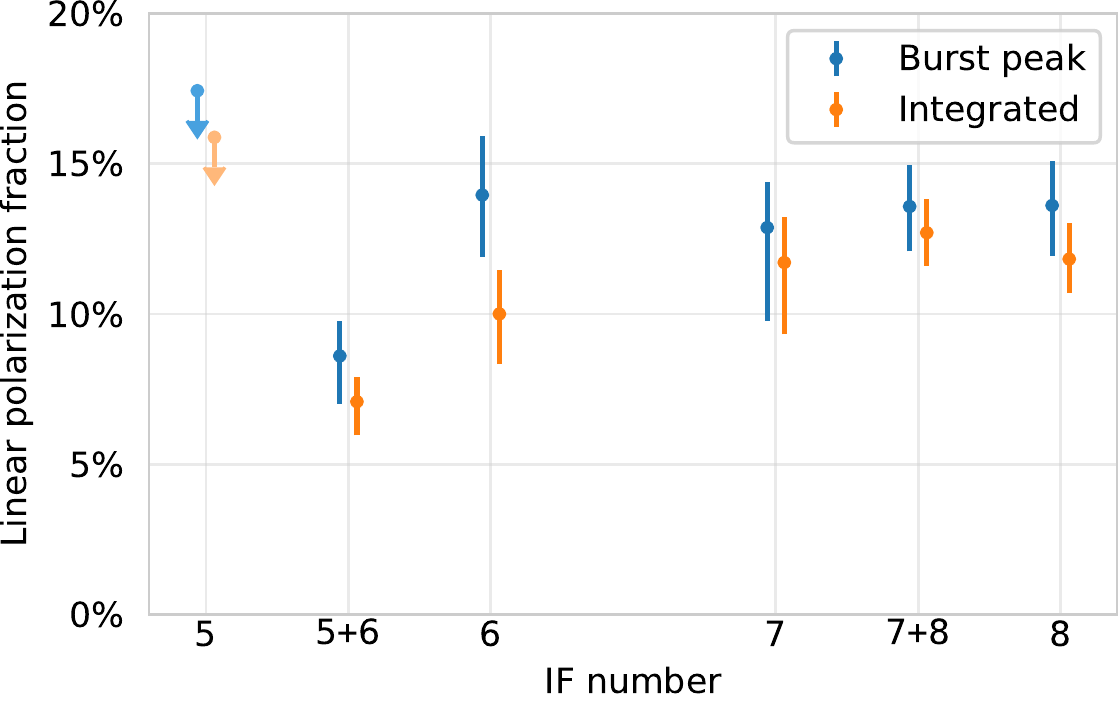}
    \hfill
    \includegraphics[width=\linewidth]{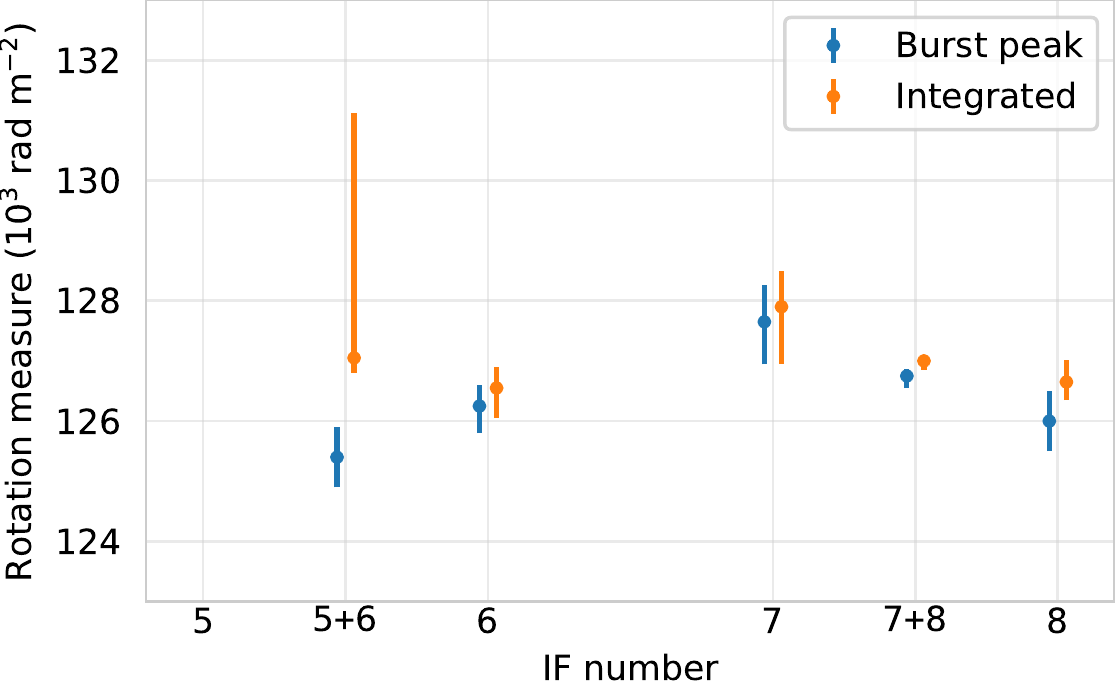}
    \caption{Linear polarization properties of the 2016 Sep 20 burst from FRB~121102: the polarization fraction and the rotation measure. Results are shown for subbands where signal is consistently detected, IFs 6, 7, and 8, and two corresponding IF pairs. Burst peak corresponds to $t=0$~ms in \autoref{f:burst_flux}. Error bars indicate 68\% statistical uncertainties; upper limits are shown at a 95\% level.}
    \label{f:rm_vals_fracs}
\end{figure}

The total intensity (Stokes I) burst profile is shown in \autoref{f:burst_flux}, separately for each of the eight subbands (IFs).  Here, times are referenced to the burst peak. IFs 1 and 2 have much worse noise and calibration properties than others: we find that they were stronger affected by radio interference. We thus drop IFs 1 and 2 from further analysis and show them in grey in \autoref{f:burst_flux}.

We show the rotation measure spectrum $|S(\mathrm{RM})|$ at $t=0$~ms in \autoref{f:rm_profile} for IFs joined into pairs, see \autoref{s:method_sd} for details. Profiles for each individual IF and for time-integrated emission are presented in appendix~\ref{a:rm_profiles} for completeness. The highest significant peaks consistently lie above $3.5\sigma$ level for IFs 6, 7, and 8, and for corresponding IF pairs. These peaks are all located around $RM \sim 1.27\cdot10^5$~rad~m$^{-2}$. Exact values and locations of maxima in these profiles are presented in \autoref{f:rm_vals_fracs}, together with their uncertainties. The lowest statistical uncertainty for the rotation measure is obtained in the 7+8 IF pair: $126750\pm150$~rad~m$^{-2}$ at the burst peak. However, we cannot completely rule out potential systematic effects, and conservatively add the inter-IF scatter to those formal errors. This leads to our final estimates of rotation measure $RM=126750\pm800$~rad~m$^{-2}$ and polarization fraction $L/I=13\pm3\%$. We check and confirm that the detection and parameter estimation of this linearly polarized emission does not strongly depend on the calibration, as expected (\autoref{s:method_sd}). Indeed, the peak in $S(\mathrm{RM})$ profiles is present even in the raw unedited data at consistent RM values, albeit with a higher noise level.

The RM profile peaks are statistically significant, but the signal is only a few times above the noise. At low S/N ratios, it is challenging to properly evaluate all potential systematic effects that could affect these measurements. As detailed above, we find the peak to stay consistent for all IFs where polarization is detected, and for different data processing setups. Despite these cross-checks, we urge everyone to treat this polarization detection with care.

Further, we test for potential time- and frequency variations of the polarization properties that could lead to smearing and decrease the polarization fraction. We repeated the same analysis at four times higher time resolution and split the band into IFs four times narrower. Despite a lower sensitivity of such setup, linear polarization would still be detectable if its fraction was close to 100\%. However, no signal is detected at any rotation measure in this case. Thus, we conclude that the 1.7-GHz burst emission is indeed polarized to a much smaller degree compared to higher frequencies \citep{michilli2018,2020arXiv200912135H}; lower polarization is not caused by temporal or frequency smearing at the probed scales.

We do not detect circular polarization at any conversion measure in any of the IFs. Sensitivity is mostly limited by the amplitude calibration uncertainty, at least close to $CM=0$: see \autoref{s:method_sd}. We put a conservative upper limit of $V/I < 15\%$ for all IFs.

\subsection{Persistent source}
\label{s:res_pers}

\begin{table*}
    \centering
    \begin{tabular}{rccccrr}
\hline
Date & Frequency & Flux density & Polarized flux density & Apparent size & \multicolumn{2}{c}{Relative position} \\
(YYYY-MM-DD) & (GHz) & ($\mathrm{\upmu Jy}$) & ($\mathrm{\upmu Jy}$) & (mas)           & RA (mas)         & Dec (mas) \\
\hline
2017 Feb 23   & 1.7   & $239 \pm 62$          &                       & $2.69 \pm 0.10$ & $-0.64 \pm 0.87$ & $0.60 \pm 0.87$ \\
2017 Mar 01   & 4.8   & $150 \pm 17$          &                       & $0.38 \pm 0.03$ & $-0.03 \pm 0.09$ & $-0.14 \pm 0.09$ \\
2017 May 31   & 1.7   & $278 \pm 54$          & $< 139$               & $3.64 \pm 0.25$ & $-0.09 \pm 0.64$ & $0.91 \pm 0.64$ \\
2017 Jun 09   & 4.8   & $138 \pm 22$          & $< \hspace{0.5em}91$                & $0.88 \pm 0.03$ & $0.12 \pm 0.14$  & $-0.01 \pm 0.14$ \\
2017 Oct 27   & 4.8   & $132 \pm 12$          & $< \hspace{0.5em}32$                & $0.29 \pm 0.02$ & $0.00 \pm 0.06$  & $0.00 \pm 0.06$ \\
2017 Nov 03   & 1.7   & $232 \pm 32$          & $< \hspace{0.5em}88$                & $2.40 \pm 0.11$ & $0.15 \pm 0.40$  & $0.13 \pm 0.40$ \\
\hline
\multirow{2}{*}{Aggregated} 
   & 1.7   & $243 \pm 19$       & $< \hspace{0.5em}88$                & $2.40 \pm 0.11$ &               &  \\
   & 4.8   & $138 \pm 8$        & $< \hspace{0.5em}32$                & $0.29 \pm 0.02$ &               &  \\
\hline
    \end{tabular}
    \caption{Properties of the persistent radio source at different frequencies and epochs. Position is given relative to the most precise measurement on 2017 Oct 27 at $4.8$ GHz. Aggregated values are weighted averages of the flux densities and relative positions, and minima of the polarized flux density upper limits and apparent size estimates. Note the significant difference in the source size between the 1.7 and 5-GHz observations. This is likely to be caused by a significant scatter broadening in the region (see text for details). The apparently significant changes in size over time for the same frequency indicate underestimated formal errors for this very faint target, rather than true size variations.}
    \label{tab:obs}
\end{table*}

\begin{figure*}
    \centering
    \begin{subfigure}[b]{0.45\textwidth}
    	\includegraphics[width=\textwidth]{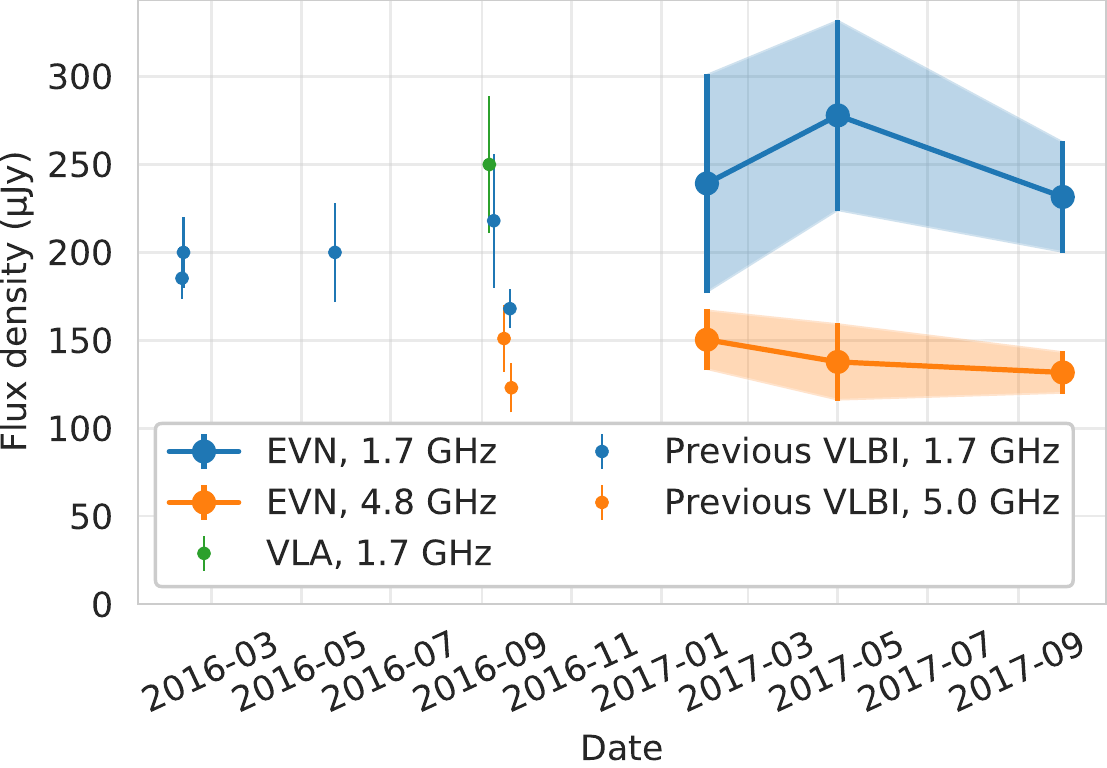}
    	\caption{Flux density measurements of the FRB 121102 persistent counterpart. Our results based on EVN observations during 2017 are shown together with previous measurements at VLA, VLBA and EVN \citep{marcote2017, chatterjee2017}. Error bars represent $1\sigma$ bounds. Variability within each frequency band is not significant and below $10\%$.}
    	\label{f:persflux}
    \end{subfigure}
	\hfill
        \begin{subfigure}[b]{0.45\textwidth}
    	\includegraphics[width=\textwidth]{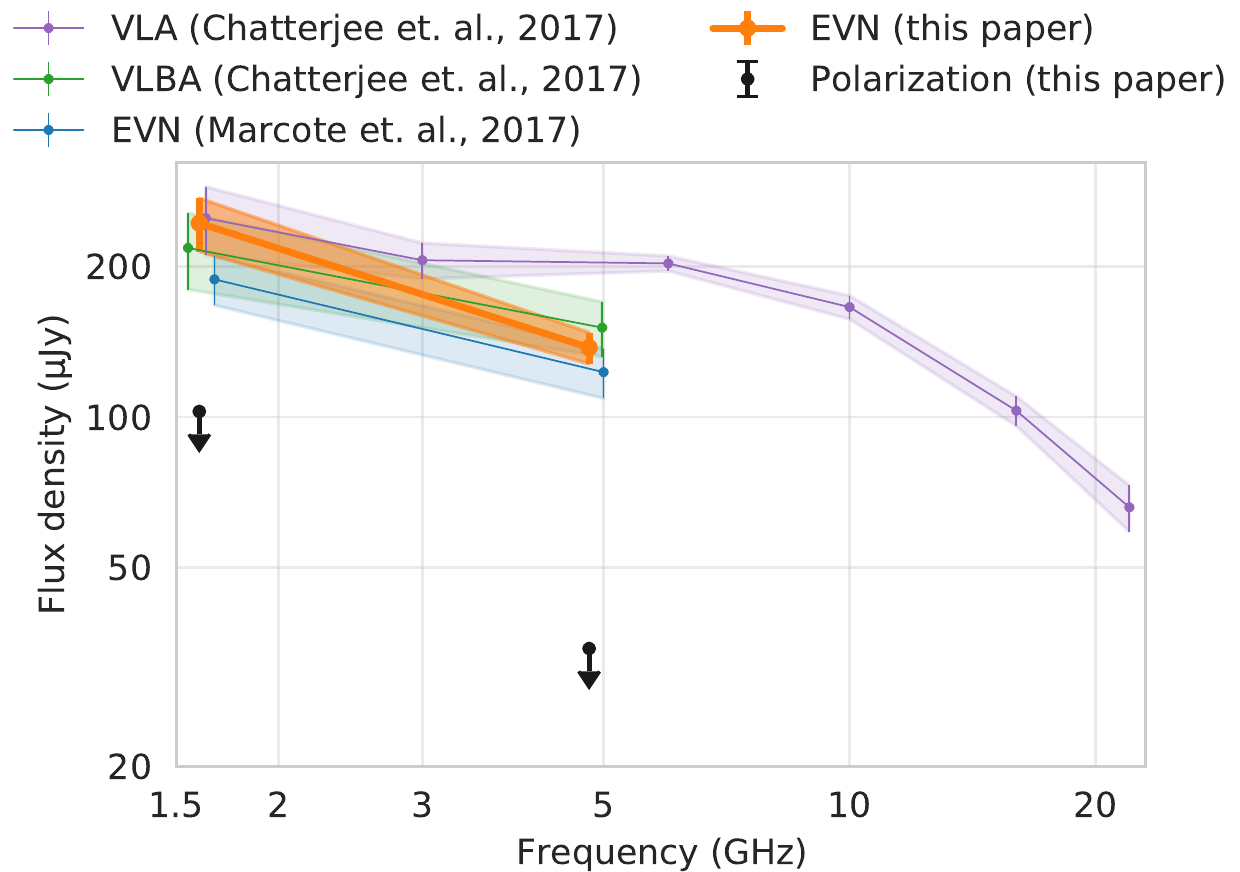}
        \caption{Spectrum of the FRB 121102 persistent radio counterpart. Includes our EVN observations in 2017, the VLA, EVN and VLBA observations in 2016 \citep{chatterjee2017}, and the EVN observations in the same year \citep{marcote2017}. Error bars represent $1\sigma$ bounds, and the polarization upper limits are at the $5\sigma$ level of the dirty map noise. Our results put the tightest constraint on the $4.8$~GHz VLBI flux density: it shows an almost $5\sigma$ difference from the VLA measurement.}
    	\label{f:persspec}
    \end{subfigure}
    \caption{Persistent source flux density measurements.}
\end{figure*}

\begin{figure*}
    \centering
    \begin{subfigure}[b]{0.45\textwidth}
        \centering
        \includegraphics[width=\textwidth]{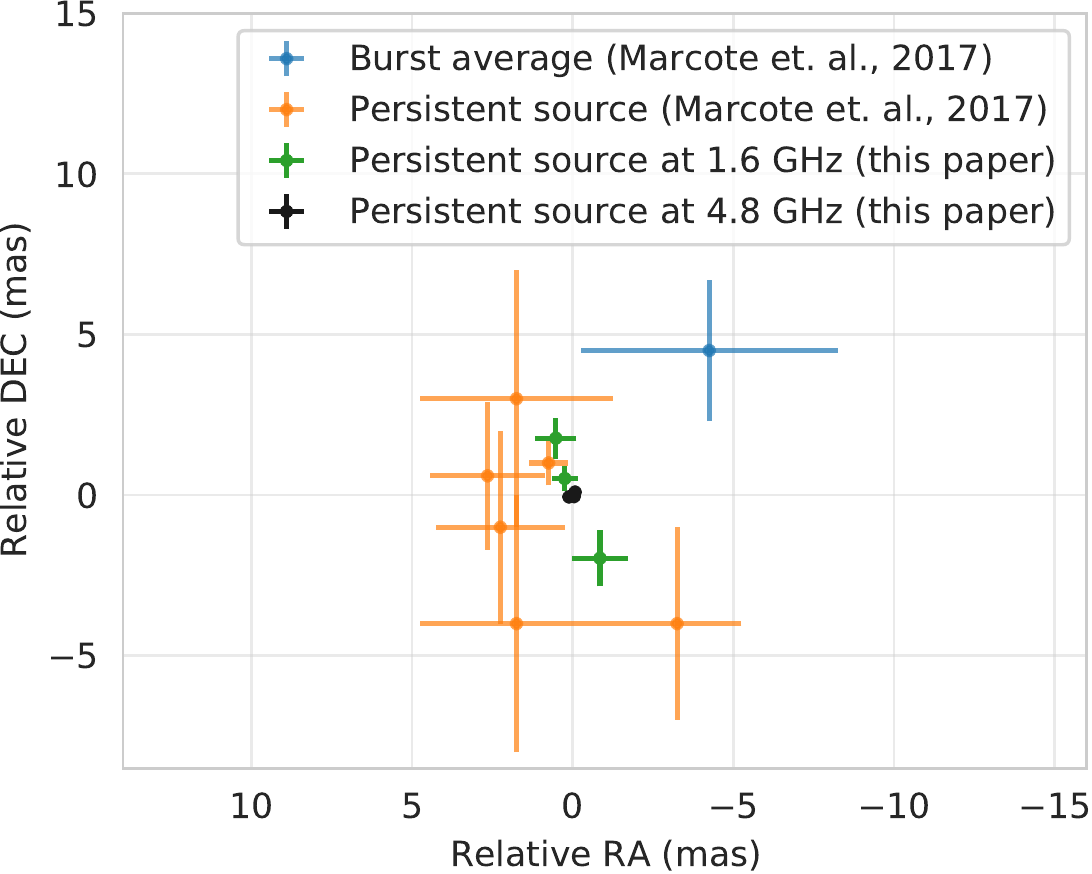}
        \caption{Comparison of our absolute coordinate measurements of the persistent counterpart with its previous measurements and with the burst position. All positions are consistent at a $3\sigma$ level.}
        \label{f:a}
    \end{subfigure}
	\hfill
    \begin{subfigure}[b]{0.45\textwidth}
        \centering
        \includegraphics[width=\textwidth]{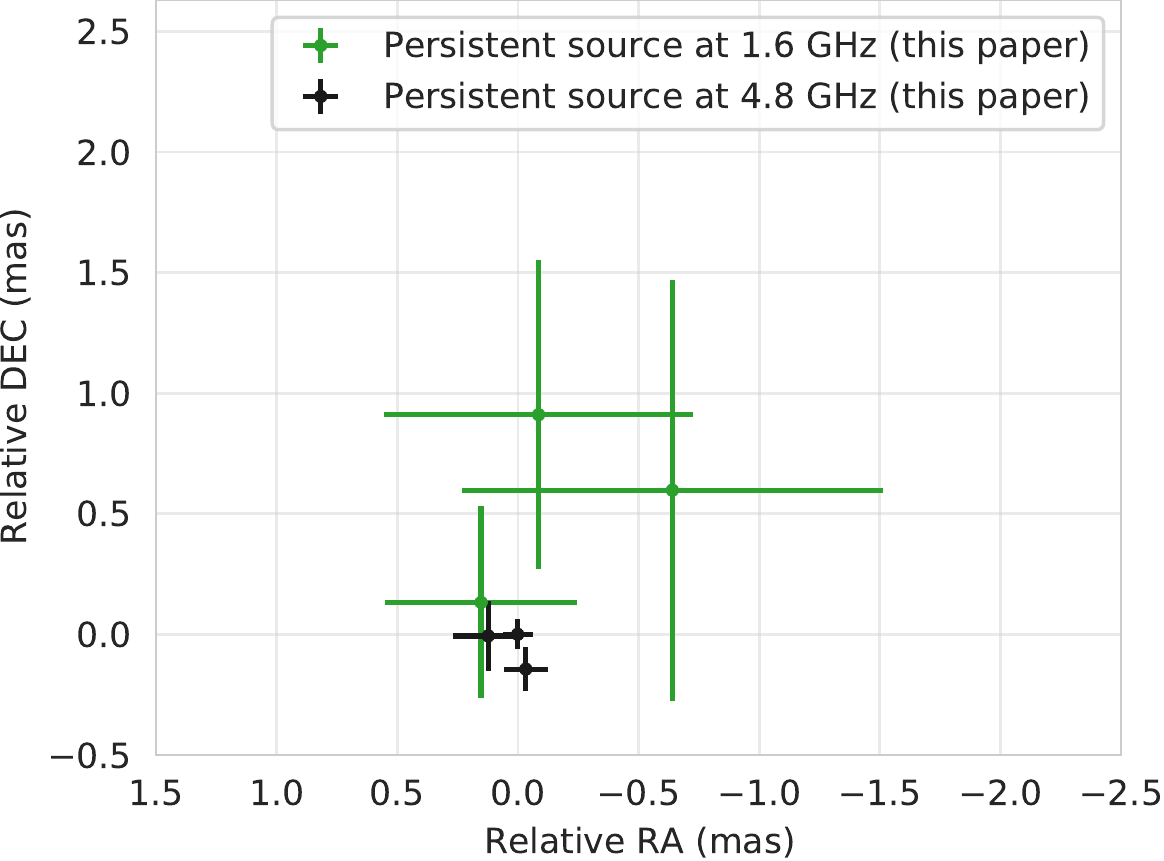}
        \caption{Our relative persistent source position measurements at 1.7 and 4.8 GHz. This uses the in-beam source as the reference (\autoref{s:method_vlbi}).}
        \label{f:b}
    \end{subfigure}
    \caption{Position of the FRB 121102 persistent radio counterpart and comparison with the burst itself. Centered on our most accurate persistent source localization at 4.8~GHz. Error bars are drawn at a $1\sigma$ level.}
    \label{fig:locations}
\end{figure*}

The persistent source associated with FRB~121102 was detected in all six EVN observations as a compact source on milliarcsecond scales. 
The better $uv$-coverage and sensitivity of these new observations allowed us to improve the localization of the source and the flux density measurements with respect to the previous results published in \citet{marcote2017}. Table~\ref{tab:obs} summarizes all analysed observations and obtained results.

The source exhibits an average flux density of $243 \pm 19\ \mathrm{\upmu Jy}$ at 1.7\,GHz and $138 \pm 8\ \mathrm{\upmu Jy}$ at $4.8$\,GHz, with variations of $\lesssim 10\%$ at both frequencies: see \autoref{f:persflux}. No significant variability is detected, thus it is appropriate to analyse the average spectrum and compare observations made at different epochs with different instruments. Comparison of our measurements with those obtained at the VLA, VLBA and EVN in 2016 \citep{chatterjee2017,marcote2017} is shown in \autoref{f:persspec}. Both $1.7$ and $4.8$\,GHz flux densities are consistent at a $2\sigma$ or better level with other VLBI observations at VLBA and EVN. However, our observations yield a significantly lower flux density at $4.8$\,GHz compared to the interpolated VLA spectrum.
We detect neither circular nor linear (at $|\mathrm{RM}| < 2\cdot10^5$~rad~m$^{-2}$) polarization from the persistent source, and thus provide conservative upper limits based on the noise level: $< 23\%$ at 4.8\,GHz, and $<36\%$ at 1.7\,GHz.

The location of the persistent source is consistent across all epochs at $4.8$\,GHz within $< 0.1\ \mathrm{mas}$, as illustrated in \autoref{fig:locations}. At $1.7$\,GHz there is a significant scatter at the level of 2\,mas --- more than $3\sigma$ of statistical uncertainties. However, we find that the scatter is caused by an imperfect correction for the ionospheric contribution. Indeed, the position of the target calculated relative to the in-beam check source ($1.8^{\rm\prime}$~apart) instead of the calibrator ($1\degr$~apart) is much more stable and consistent with statistical errors, even at the $1\sigma$ level. This is apparent from comparison of the left- and right-hand panels of \autoref{fig:locations}. Thus, we use the position relative to the calibrator only to provide absolute astrometry of the target. The resulting absolute position is consistent with that previously reported by \citet{marcote2017} for the persistent source and for a burst:
\begin{alignat}{2}
     \alpha\ (\mathrm{J2000}) & {}= 5^{\rm h}31^{\rm m}58.7016^{\rm s} & \pm 0.8 {\rm ~mas} \\
     \delta\ (\mathrm{J2000}) & {}= 33^{\circ}8^{\prime}52.5491^{\prime\prime} & \pm 0.8 {\rm ~mas}
\end{alignat}
at the reference epoch of 2017.5. The errors are currently dominated by the calibrator position uncertainty, and not by our relative measurements.

Our relative position measurements are consistent across all three epochs and at both frequencies. This lets us constrain the apparent proper motion to $\lesssim 0.1\ \mathrm{mas\ yr^{-1}}$ ($\lesssim 0.3c$ at the host galaxy distance, \cite{chatterjee2017}), and the distance between emission regions at 1.7 and 4.8\,GHz to $\lesssim 0.4\ \mathrm{mas}$. The upper limits are about 20 times below those reported in \cite{marcote2017} due to a better coverage in both the \textit{uv} plane and time, especially at 4.8\,GHz. The lack of proper motion agrees with the extragalactic origin of the persistent source established in earlier studies.

Fitting a single Gaussian to the data shows that the persistent source exhibits a significant apparent size at both bands, with FWHM of $\gtrsim 2.4$\,mas at 1.7\,GHz and $\gtrsim 0.3$\,mas at $4.8$\,GHz; for comparison, the synthesized beam size of the most constraining observation at $4.8$\,GHz is $1.2 \times 0.9$\,mas$^2$. However, this apparent size is most likely dominated by scatter broadening \citep[as previously noted by][]{marcote2017}. Indeed, the obtained sizes at 1.7 and 4.8\,GHz follow the expected $\nu^{-2}$ relation, and they are similar for both the persistent source and the in-beam calibrator ($1^{\prime}$ apart). We also note that the measured size is consistent with that of the bursts \citep{marcote2017}, which can only arise from an extremely compact region (due to the millisecond-duration of the bursts). This means that we can only derive an upper bound on the intrinsic source size based on the Gaussian FWHM, and this bound is $\lesssim 0.29$\,mas. Given the angular diameter distance to FRB~121102 of $\approx 683
~\mathrm{Mpc}$ \citep{tendulkar2017}, the obtained source size represents a physical size of $\lesssim 1.0~\mathrm{pc}$.

\section{Discussion}
\label{s:discuss}

\begin{figure}
    \includegraphics[width=\linewidth]{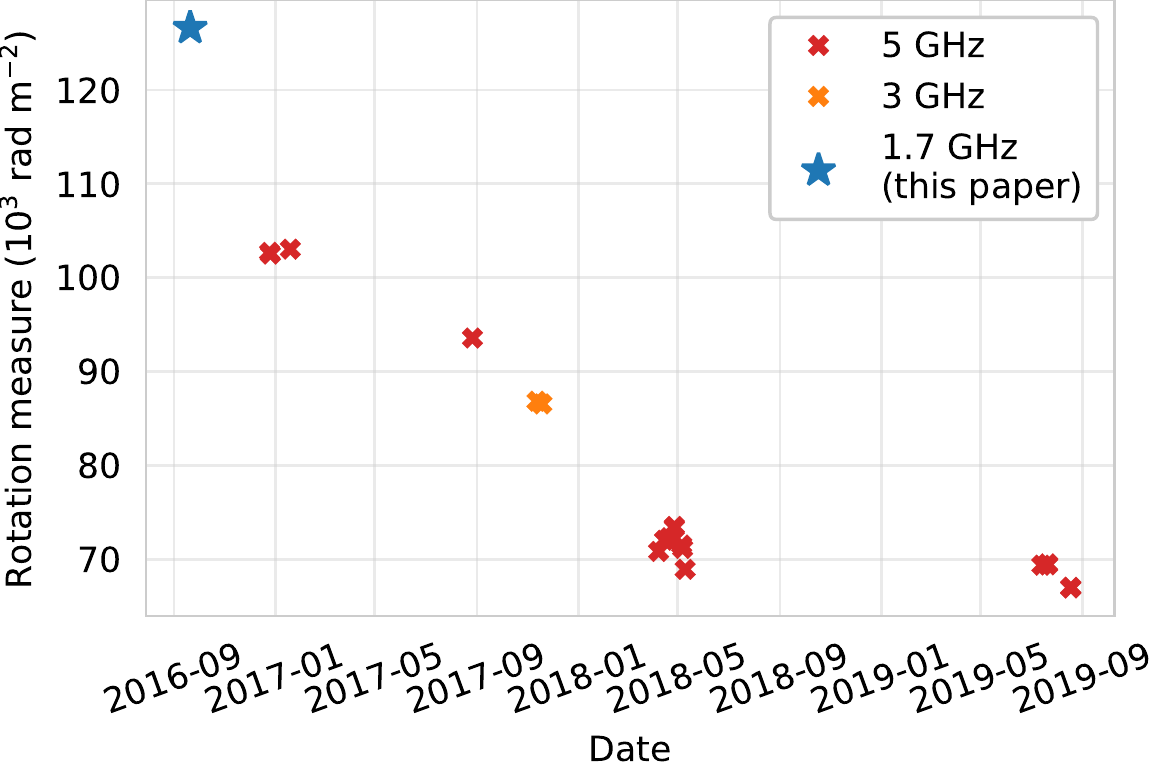}
	\caption{Comparison of our rotation measure estimates at 1.6~GHz to those previously obtained at higher frequencies \citep{michilli2018,2020arXiv200912135H}. Formal errors of these measurements are too small to be visible here. All RM values are specified in the observer frame of reference.}
	\label{f:rm_evolution}
\end{figure}

\begin{figure}
    \includegraphics[width=\linewidth]{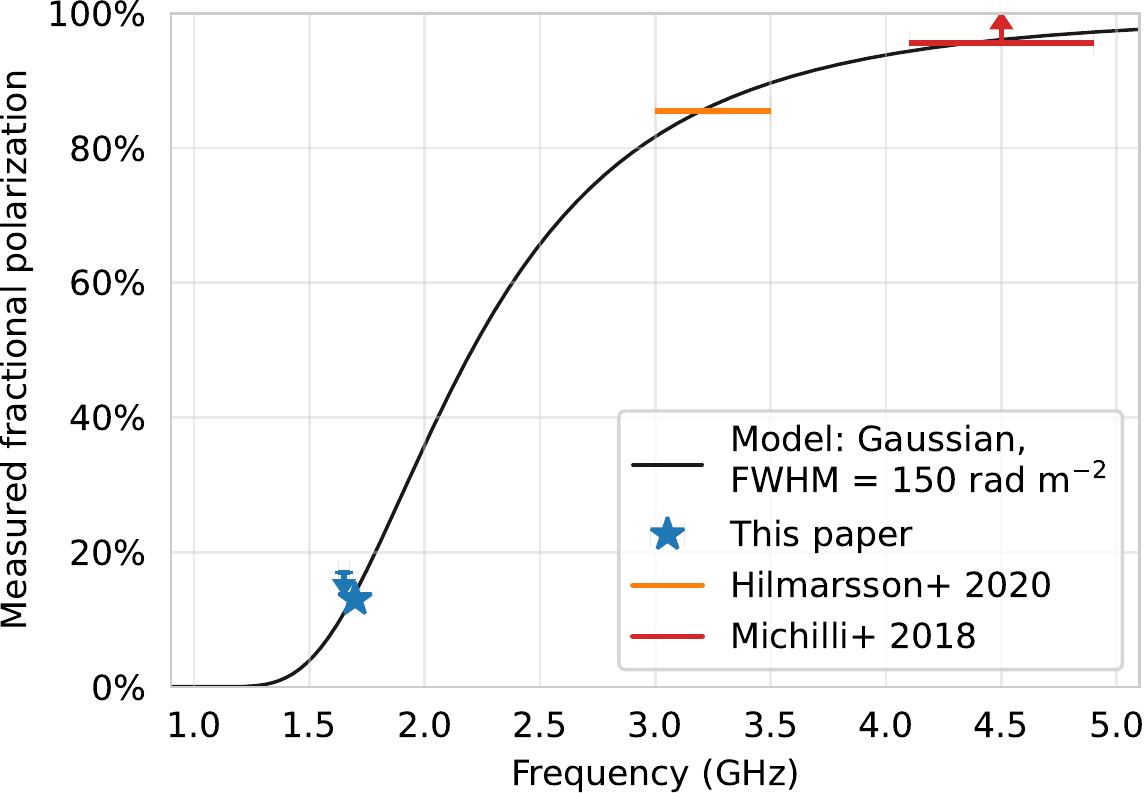}
	\caption{Measured fractional polarization at different frequencies. The emission is assumed to be intrinsically 100\% polarized with effective width in the rotation measure space of 150~rad\,m$^{-2}$ (Gaussian profile). This phenomenological model is consistent with our measurements at 1.7~GHz and with higher 3-5~GHz measurements \citep{michilli2018,2020arXiv200912135H}.}
	\label{f:fracpol_freq}
\end{figure}

To put our results into the broader picture of FRB~121102, we compare them with observations performed at different radio frequencies. No simultaneous detections of the bright burst on 2016 Sep 20 have been reported; thus, the only way to evaluate polarization properties across different radio frequencies is to compare measurements at different epochs. \autoref{f:rm_evolution} illustrates that our estimate of the burst Faraday rotation measure at 1.7\,GHz, $\mathrm{RM}=1.27\cdot10^5$\,rad\,m$^{-2}$, is the highest measured for this source to date. This value is qualitatively consistent with the decreasing trend observed at higher frequencies (\autoref{f:rm_evolution}) originally reported in \cite{michilli2018}. This can be considered an additional argument for the polarization detection being reliable at 1.7~GHz despite a low S/N ratio detailed in \autoref{s:res_burst}. Such consistency also indicates that the emission at both frequencies propagates through the same general region of the environment, and is colocated.

It is not yet clear what the complete shape of the rotation measure variability curve looks like. \autoref{f:rm_evolution} may suggest that our 1.7~GHz measurement lies significantly above the trend extrapolated from later higher-frequency observations. This difference, if actually present, can be explained in several ways. We believe temporal variability to be the most likely explanation: that is, the rotation measure is the same at all frequencies at any given moment, but varies rapidly and irregularly at the probed timescales. Another possibility is that emission at 1.7~GHz and 5~GHz passes through slightly different regions of space, thus observed rotation measures depend on frequency. Finally, it is possible for emission at different frequencies passing through the same medium to experience different Faraday rotation. This nonlinearity would require either extremely dense ($n_e \sim 10^9$~cm$^{3}$) or extremely magnetized ($B \sim 100$~G) plasma: the plasma or cyclotron frequency have to be on the order of 1~GHz. We find this scenario physically unlikely; see also \cite{michilli2018} for relevant considerations.

Our analysis in \autoref{s:res_burst} shows that the burst emission at 1.7\,GHz is linearly polarized at a $15\%$ level. This is much lower than almost completely polarized 5-GHz bursts \citep{michilli2018,gajjar2018} and $\approx 85\%$ polarized 3~GHz bursts \citep{2020arXiv200912135H}. There are various possible explanations, including linear-to-circular conversion in magnetized plasma \citep{2019MNRAS.485L..78V,2019ApJ...876...74G} or some kind of depolarization at lower frequencies. A recent study hasn't detected any burst polarization at 1 to 1.5 GHz in 2019 \citep{Li2021BimodalBurst}. Proper comparisons of those results with ours would require quantified upper limits; a potential difference can be caused by temporal variations over the years or by depolarizing effects we discuss below. We detect no circular polarization and put an upper limit of $\lesssim 15\%$, which rules out conversion as the dominant scenario. We also test for temporal depolarization and do not detect any evidence for this, as explained in \autoref{s:res_burst}.

Spatial depolarization remains a plausible hypothesis. Following \cite{2005A&A...441.1217B}, we estimate that the emission Faraday width of $\Delta \mathrm{RM} = 150$\,rad\,m$^{-2}$ is enough to reduce the measured polarization fraction from 100\% to 15\% at 1.7\,GHz. Remarkably, this $\Delta \mathrm{RM}$ value is consistent with measurements at all three frequency bands, 1.7, 3, and 5 GHz. To perform a direct quantitative comparison, we assume a Gaussian Faraday profile with FWHM $= 150$\,rad\,m$^{-2}$. We extract burst polarization fractions at higher frequencies from corresponding papers. \cite{michilli2018} puts the lower limit of $L/I > 96\%$ from 4.1 to 4.9 GHz. The lowest frequency burst in \cite{2020arXiv200912135H} is burst \#6 at $3-3.5$~GHz; it is $\approx85\%$ polarized, according to their Figure~1. We illustrate these measurements together with the simulated Gaussian outcomes in \autoref{f:fracpol_freq}. Clearly, such a simplistic model already fits the observed behaviour very closely. Note that it is possible for a burst to be apparently less polarized at the corresponding frequencies due to intrinsic or instrumental depolarization. This does in fact happen: see other bursts in \cite{2020arXiv200912135H}. We specifically only consider the highest observed polarization at each band for the comparison above.

The Faraday width of $\Delta \mathrm{RM} = 150$\,rad\,m$^{-2}$ constitutes just $0.1$\% of the total burst Faraday rotation. This width could arise due to minor non-uniformities in the Faraday screen: different rays experience different screen depths, and the resulting $\mathrm{RM}$ varies across the source. Another explanation could be that a small part of the Faraday rotation is gained in the emitting region itself. We believe that environment non-uniformities constitute a more plausible explanation. Indeed, the Faraday rotating region co-located with the burst source itself presents various challenges: extremely high electromagnetic fields required to generate the FRB are not consistent with continuous existence of thermal plasma. Sensitive polarization observations of multiple bursts covering a wide frequency range, from 1 to 5~GHz, would help to test and distinguish these scenarios.

Our observational results related to the faint persistent radio counterpart of FRB~121102 significantly expand on what was known about this source before \citep[e.g.][]{marcote2017}. We find that the emission at both 1.7 and 5\,GHz comes from effectively the same region: the separation between the centroids projected to the sky plane is below $0.3$\,pc. This emitting region does not noticeably move on timescales of years with the apparent proper motion below $0.3c$. The upper-limit on the source size of $< 0.3~\mathrm{mas}$ ($< 1.0$\,pc) at 4.8\,GHz additionally constrains the possible source expansion rate. Assuming a $\sim 10$--50\,yr old wind nebula \citep{margalit2018}, it expands no faster than $(1\text{--}5) \cdot10^4\ \mathrm{km\ s^{-1}}$. We note that these limits are close to the typical ejecta speed of $\sim 10^4~\mathrm{km\ s^{-1}}$ for hydrogen-poor supernovae.

The flux density of the faint persistent source is stable as well: we show that possible variations across a year are $< 10$\%. This already somewhat constrains potential models of the persistent radio counterpart \citep[see][for a review of models]{2019PhR...821....1P}: e.g., an expanding supernova would show a decaying luminosity trend on year timescales \citep[see e.g.][]{margalit2018}; a typical AGN would have significant frequency-dependent position differences (core shift), or bright features moving along the jet that lead to a position jitter, or a general flux variability. We note that some models \citep[see e.g.][]{beloborodov2017} predicted an increase of the persistent emission on scales of half a year after episodes of FRB activity. While some of the activity episodes have been observed during these years (e.g., August 2016 that led to the localization of the bursts), no significant increase in the persistent emission is reported. Our upper limit on the persistent source linear polarization of $< 36\%$ at 1.7\,GHz is compatible to measurements for the burst itself ($15\%$). However, at 5\,GHz the bursts were shown to be almost completely polarized \citep{michilli2018}, while the persistent source still shows no significant polarization with the upper limit of $< 25\%$. This rules out the possibility of bursts and the persistent emission having the same nature, i.e. that the persistent emission is due to the sum of frequent, low-level burst activity.

The comparison of the persistent source spectra shown in \autoref{f:persspec} illustrates that the compact pc-scale (VLBI) emission has a steeper spectrum compared to that of a more extended (VLA) emission. This is an unusual effect: typically, the spectrum of extended emission regions is steeper. The observed behaviour may in principle be explained by an instrumental effect: VLBI observations probe smaller spatial scales at higher frequencies and may effectively resolve out some flux at $4.8$\,GHz despite fully detecting emission from the same region at 1.7\,GHz. A physical explanation of the spectra difference could be that the persistent source is variable intrinsically or due to propagation effects, and we just do not have enough sensitivity to detect the variations. However, this is unlikely as our VLBI measurements are consistent with earlier ones and together cover a time range of more than 1.5 years at 1.7\,GHz, and a complete year at 4.8\,GHz. More observations, especially simultaneously at different angular scales, could test whether the difference is due to variability. If the difference between the compact and extended spectra is indeed astrophysical, it constitutes an additional challenge to models that explain both FRB~121101 bursts and its environment.

\section*{Acknowledgements}

The European VLBI Network is a joint facility of independent European, African, Asian, and North American radio astronomy institutes.
This work was also based on simultaneous EVN and PSRIX data recording observations with the 100-m telescope of the Max-Planck-Institut f\"ur Radioastronomie at Effelsberg, and we thank the local staff for this arrangement. The Astronomical Image Processing System (AIPS) is a software package produced and maintained by NRAO. B.M. acknowledges support from the Spanish Ministerio de Econom\'ia y Competitividad (MINECO) under grant AYA2016-76012-C3-1-P and from the Spanish Ministerio de Ciencia e Innovaci\'on under grants PID2019-105510GB-C31 and CEX2019-000918-M of ICCUB (Unidad de Excelencia ``Mar\'ia de Maeztu'' 2020-2023).
 J.W.T.H. acknowledges funding from an NWO Vici grant (``AstroFlash''). LGS is a Lise Meitner research group leader and acknowledges funding from the Max Planck Society.
 
\section*{Data Availability}

The data underlying this article are available in the EVN Data Archive at JIVE at \url{https://www.jive.eu/select-experiment}, and can be accessed with project codes EP103, RP026.

\appendix

\section{Faraday rotation profiles}
\label{a:rm_profiles}

\begin{figure}
    \begin{subfigure}[b]{\linewidth}
    	\includegraphics[width=\linewidth]{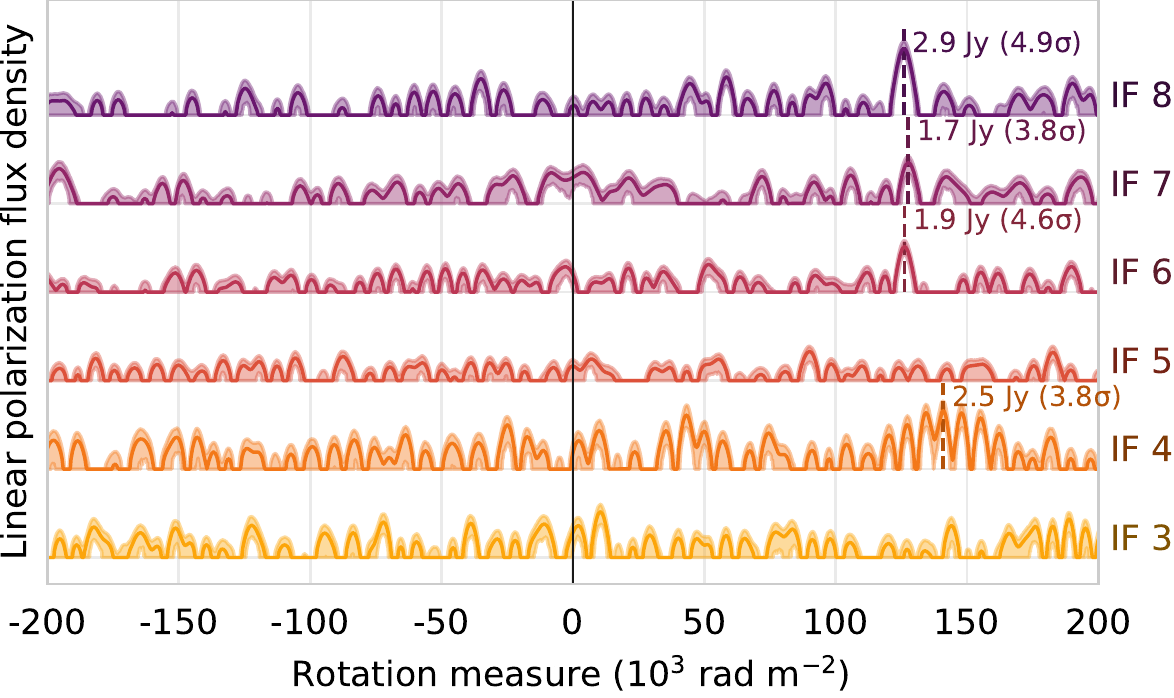}
    	\caption{Each IF is shown separately instead of aggregating into pairs.}
    \end{subfigure}
    \begin{subfigure}[b]{\linewidth}
    	\includegraphics[width=\linewidth]{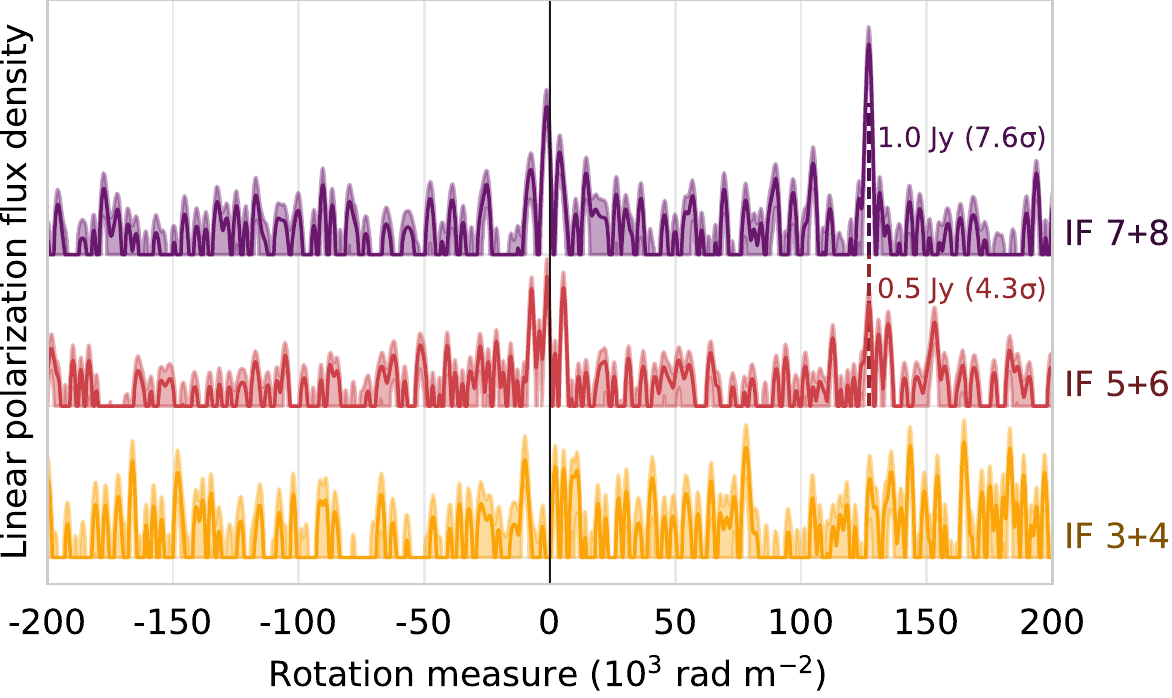}
    	\caption{Emission is integrated over the whole burst duration, from -1 to 1~ms (see \autoref{f:burst_flux}).}
    \end{subfigure}
    \begin{subfigure}[b]{\linewidth}
    	\includegraphics[width=\linewidth]{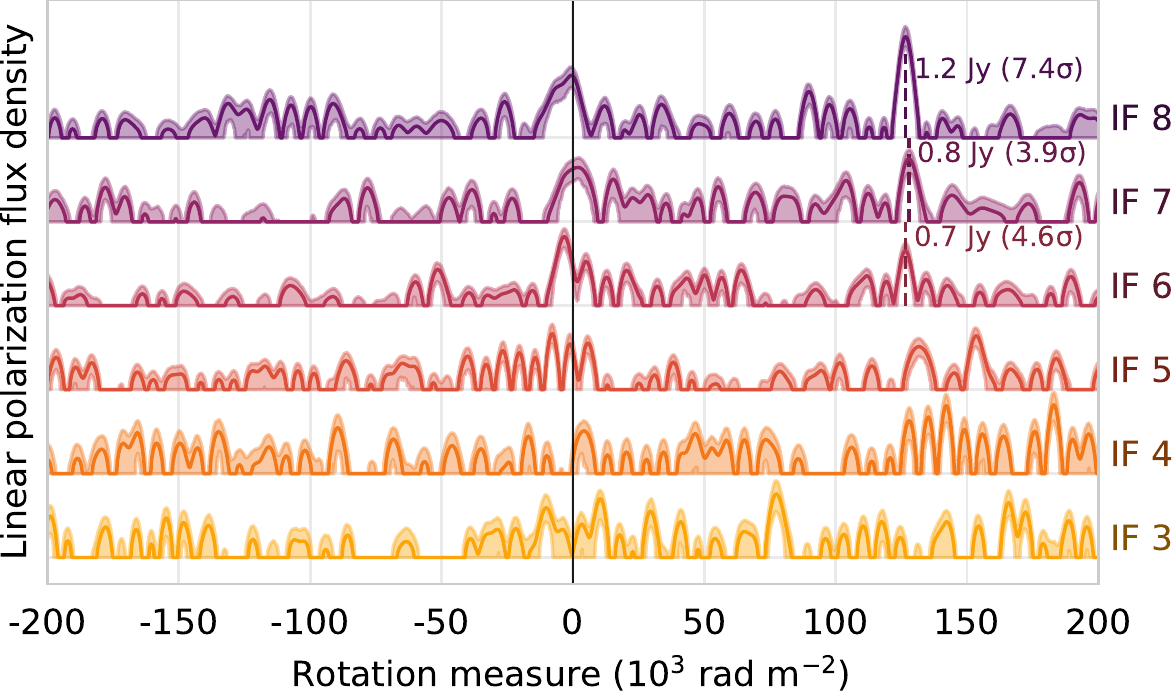}
    	\caption{Each IF is shown separately, and emission is integrated over the whole burst duration.}
    \end{subfigure}
	\caption{Faraday rotation measure profiles for the 1.7~GHz burst. Same as \autoref{f:rm_profile}, but with different options.}
    \label{f:rm_profiles_other}
\end{figure}

Rotation measure profiles shown in \autoref{f:rm_profile} are calculated for the burst peak and aggregated over consecutive pairs of IFs. As described in \autoref{s:method_sd}, we perform a consistency check by applying the same process to each IF individually, and to emission temporally integrated over the whole burst. Profiles computed with different options are presented in \autoref{f:rm_profiles_other} that extends \autoref{f:rm_profile}.

When integrating over the whole burst (\autoref{f:rm_profiles_other} b and c), spurious peaks appear in the region close to $RM=0$~rad~m$^{-2}$. We attribute these peaks to uncalibrated instrumental polarization effects: their scales in the rotation measure space agrees with expectations, see \autoref{s:method_sd}. Outside this region, we see the highest peaks above the $3.5\sigma$ level consistently located at $(125-128)\cdot10^3$~rad~m$^{-2}$, with a single exception of IF~4 in \autoref{f:rm_profiles_other} (a).

We thus confirm that linearly polarized emission is reliably detected at the higher half of our band. Rotation measure and polarization fraction estimates stay consistent for different data aggregation choices.

%%%%%%%%%%%%%%%%%%%%%%%%%%%%%%%%%%%%%%%%%%%%%%%%%%

\bibliographystyle{mnras}
\bibliography{main}

% Don't change these lines
\bsp	% typesetting comment
\label{lastpage}
\end{document}